\documentclass[12pt]{article}
\usepackage{a4}
\usepackage{amssymb}
\usepackage{cite}
\def\Bbb{\mathbb}
\def\BZ{\mbox{$\Bbb Z$}} 
\def\BC{\mbox{$\Bbb C$}} \def\BP{\mbox{$\Bbb P$}}

\catcode`@=11 \@addtoreset{equation}{section} \catcode`@=12

\begin{document}
\begin{titlepage}
\noindent
{\tt TIFR/TH/00-25} \hfill
{\tt hep-th/0007075}\\
{\tt IMSc/2000/06/22}~~~
\hfill July, 2000
\vfill
\begin{center}
{\Large \bf On D-branes from Gauged Linear Sigma Models} \\[1cm]
Suresh Govindarajan\footnote{Email: suresh@chaos.iitm.ernet.in}\\
{\em Department of Physics, Indian Institute of Technology, Madras,\\
Chennai 600 036, India\\[10pt]}
T. Jayaraman\footnote{Email: jayaram@imsc.ernet.in}\\
{\em The Institute of Mathematical Sciences, \\ Chennai 600 113, India\\
[10pt]}
Tapobrata Sarkar\footnote{Email: tapo@theory.tifr.res.in}\\
{\em Department of Theoretical Physics, \\ Tata Institute of 
Fundamental Research, \\ Homi Bhabha Road, Mumbai 400 005, India}\\ 
\end{center}
%\vspace{2cm}
\vfill
\begin{abstract}
We study both A-type and B-type 
D-branes in the gauged  linear sigma model by considering
worldsheets with boundary. The boundary conditions on the matter and
vector multiplet fields are first considered in the 
large-volume phase/non-linear sigma model limit of the corresponding 
Calabi-Yau manifold, where we 
find that we need to add a contact term on the boundary. 
These considerations enable to us to derive the
boundary conditions in the full gauged linear sigma model, including the
addition of the appropriate boundary contact terms, such that these
boundary conditions have the correct non-linear sigma model limit. 
Most of the analysis is for the case of Calabi-Yau manifolds with
one K\"ahler modulus (including those corresponding to hypersurfaces
in weighted projectve space), though we comment on possible generalisations.
\end{abstract}
\vfill
\end{titlepage}

\section{Introduction}

The improved understanding of non-perturbative aspects of string theory in
recent years has shown that all five (perturbative) superstrings appear to
be different corners in the moduli space of a single theory\cite{schwarz}.
A consequence of this is that while the existence of a perturbative string
theory description at these corners singles out strings as fundamental
objects, at generic points in this moduli space, there is a certain
democracy among all objects, fundamental as well as solitonic, as seen in
the perturbative string theory. Thus, the heterotic string appears as a
soliton (D1-brane) in the type I theory. A more general analysis indicates
that an object which is a soliton at one point in the moduli space can
become a fundamental excitation at another point in the moduli
space\cite{hull}. 

Another question of interest is the nature of spacetime at short
distances. It turns out that the answer is related to the kind of probe
which is used. Given the democracy among all objects one can probe
spacetime using both fundamental strings as well as solitons such as
D-branes. Fundamental strings probe objects which are of the string scale
$l_s$ while D-branes in perturbative string theory probe much shorter
scales $(g_s)^a l_s$, where $g_s$ is the string coupling constant and $a$
is a positive constant\cite{scales}. From earlier studies of closed
strings, it is known that strings can propagate in apparently singular
spaces such as orbifolds. D-brane probes see these space-time geometries in
a manner different from that of closed string theories. For example, it
was shown that for D0-brane probes of the orbifold $\BC^3/\Gamma$ (where
$\Gamma$ is a discrete subgroup of $SU(3)$), the non-geometric phases seen
by the closed string are projected out\cite{dgm}. 

While much is known about the nature of fundamental strings probing
various space-time geometries, our understanding in the case of D-brane
probes is in a much more primitive state, apart from the cases of flat
space and toroidal and orbifold backgrounds.  In the last couple of years,
there has been considerable progress in understanding
D-branes in the context of string compactification on Calabi-Yau
threefolds\cite{ooy,RS,quintic,doug1,stt,boundarylg,dgepner,dfr,
ketal,HV,HIV,kklm2,doug-diac,others}. 
Unlike the case of toroidal compactifications, these correspond to fewer
unbroken supersymmetries and thus fewer constraints follow. For example,
the BPS conditions leave open the possibility of having lines of marginal
stability in the moduli space, where a D-brane can decay.  A D-brane which
fills the non-compact spacetime while also wrapping some cycle of the CY
three-fold can possess a non-trivial superpotential in its worldvolume
gauge theory.  It is of interest to derive this superpotential and its
dependence on closed string moduli. 

D-branes on CY manifolds fall into two distinct categories: A-type branes
are those which wrap special Lagrangian submanifolds while B-type branes
wrap holomorphic submanifolds of the CY manifold. In the worldsheet
description\cite{ooy}, A-type branes and B-type branes differ in the
worldsheet supersymmetry that they preserve. In the open-string channel,
A-type branes are compatible with the topological theory obtained with the
A-twist and B-type branes are compatible with the B-twist. In the
closed-string channel, the roles are reversed due to a change of sign in
the boundary conditions on the $U(1)$ currents of the $(2,2)$ worldsheet
supersymmetry algebra. In the closed-string case\cite{topwitten},
correlation functions of the observables in the topological A-model are
independent of complex structure moduli (of the CY) while those in the
topological B-model are independent of the K\"ahler moduli. The {\it
modified geometric hypothesis} proposed in \cite{quintic,doug1} is in a
sense the open-string version of this statement. Based on this, one
(loosely speaking) expects the lines of marginal stability of A-branes and
the superpotential of B-branes to be independent of K\"ahler moduli and
thus calculable in the large volume limit where classical geometry can be
applied. (See \cite{doug1} for a more detailed and careful discussion.)

Tests of the {\it modified geometric hypothesis} as well as the extended
version of mirror symmetry that includes D-branes and their world-volume
theories will need a worldsheet description of CY manifolds where both
K\"ahler and complex moduli have simple realisations. The gauged linear
sigma model (GLSM) is a suitable worldsheet description in this regard. 
As shown by Witten, this model has several phases of which the Calabi-Yau
phase is one. Thus, the enlarged K\"ahler cone which is required by mirror
symmetry naturally fits into the setup\cite{agm}.  The price one pays for
this choice is that conformal invariance on the worldsheet is obtained
only at the infrared fixed point of the GLSM. 

One of the advantages of the GSLM in the closed string case is the fact
that it unifies the different techniques that are preferred in different
regions of K\"ahler moduli space. In the Calabi-Yau phase or the large
volume phase, the GLSM description tends to the non-linear sigma model
description of strings moving on CY manifolds. In a Landau-Ginzburg phase,
the description would be in terms of N=2 supersymmetric Landau-Ginzburg
theory. In particular cases, the description at this point in the moduli
space is even more explicit when the LG theory is equivalent to the tensor
product of a set of N=2 minimal model conformal field theories. One may
hope to see a similar situation in the case of D-branes. 

In this paper, as a first step towards the eventual goal described
earlier, we study the GLSM with $(2,2)$ supersymmetry on worldsheets with
boundary. Due to the presence of a boundary, one has to specify boundary
conditions on the various fields in the GLSM such that the appropriate
linear combination of supersymmetry is preserved. In order that these
boundary conditions correspond to D-branes wrapped around various cycles of
the Calabi-Yau manifold, we first construct the boundary conditions in the
nonlinear sigma model limit of the GLSM and look for boundary conditions
in the GLSM which reduce to sensible ones in the NLSM limit. We also find
the need to introduce a boundary (contact) term in order to obtain
consistent boundary conditions. This contact term vanishes when the
theta-term in the GLSM is turned off and and its presence is justified by
considering the NLSM limit. 

The organization of the paper is as follows: In section 2 we begin by
reviewing the $d=2$, $N=2$ supersymmetric GLSM for closed strings
following \cite{wittenphases}. We then add a boundary and compute the
boundary terms generated in computing the equations of motion and the
variations of the action under supersymmetry.  We also review and extend
the work of \cite{stt,boundarylg}\ describing boundary conditions for
$d=2,N=2$ supersymmetric Landau-Ginzburg (LG) models of conformal field
theories: this will be useful in understanding the boundary conditions in
LG phases of Calabi-Yau compactifications. We close with a few words about
the justification for using the GLSM; in particular we discuss the
topological twisting of the GLSM with boundary. In section 3 we construct
boundary conditions describing branes on supersymmetric cycles in the $e^2
\rightarrow \infty$ of the GLSM, as a guide to understanding the physical
meaning of boundary conditions at finite $e^2$; along the way we will find
certain boundary terms that we must add for consistency.  In section 4 we
finally construct and identify boundary conditions at finite $e^2$. We
also discuss the significance of these boundary conditions. In section 5
we present our conclusions. 

Parts of this work have been reported earlier
elsewhere\cite{thesis,talks}.  While readying this work for publication,
other papers \cite{HV,HIV,doug-diac} have also appeared that, in part,
study D-branes in the GLSM approach. While \cite{doug-diac} use a
combination of the GLSM and world-volume techniques , \cite{HIV} is closer
in spirit to the techniques of this paper and the results of sec. 6 of
their paper have some overlap with sec. 4 of this work.

\section{The Gauged Linear Sigma Model}

In type II string theories compactified on Calabi-Yau threefolds, the
moduli space of K\"ahler classes includes regions where the size of the
manifold is of order $l_s$.  Often the CFTs appear non-geometric and are
better described via Landau-Ginzburg orbifolds \cite{wittenphases,ag}; or
they may mediate smooth passage to Calabi-Yau manifolds with different
topology \cite{agm}. We are particularly interested in studying the
physics of D-brane probes as one moves through large distances in the
moduli space of such compactifications, across phases or towards singular
compactifications.  Unfortunately even in the geometric phases of these
models the Calabi-Yau metric is not known, and physical objects (such as
vertex operator correlation functions) receive corrections from worldsheet
instantons. At best there are a few points in the moduli space where the
conformal field theory is well understood: in particular at large-radius
limits, and exactly solvable Gepner points, the latter being deep in the
Landau-Ginzburg region. 

The technique introduced in \cite{wittenphases} to study motion between
these regions was to write a 2d supersymmetric field theory with a known
UV Lagrangian whose infrared fixed point is believed to be a Calabi-Yau
compactification. (In fact, this technique was an important part of the
development of the above picture.) This model is simply a $d=2$, $N=2$
supersymmetric gauge theory with some number of vector multiplets and some
number of charged chiral multiplets, and is called the gauged linear sigma
model (GLSM). This is much in the spirit of using Landau-Ginzburg models
as UV Lagrangian descriptions of minimal models: indeed, Landau-Ginzburg
orbifolds appear in ``non-geometric'' phases of the GLSM. 

\subsection{GLSM for closed strings}

For ease of reference, we will review the lagrangian and
supersymmetries of the GLSM following \cite{wittenphases}.  We work in
Minkowski space with the metric $(-,+)$.  We are interested in describing
compactifications of string theory with eight supercharges; the worldsheet
conformal field theory must then have $N=(2,2)$ superconformal symmetry. 
We expect that a nonconformal theory with such an infra-red fixed
point should have $N=2$ supersymmetry as well. 

Our candidate theory can be obtained by dimensional reduction from
$d=4,N=1$ abelian gauge theory with chiral multiplets.  It contains $s$
$U(1)$ vector multiplets, described by the vector superfields $V_a
(a=1,\cdots, s)$ and $k$ chiral multiplets described by the chiral
superfields $\Phi_i (i= 1,\cdots,k)$. Written in components, the vector
multiplet consists of the vector fields $v^a_{\alpha} (\alpha=0,1)$, the
complex scalar field $\sigma^a$, complex chiral fermions
$\lambda_{\pm}^a$, and the real auxiliary field $D^a$.  The chiral
multiplet consists of a complex scalar $\phi_i$, complex chiral fermions
$\psi_{\pm i}$, and a complex auxiliary scalar field $F_i$. They are
charged under the $U(1)$s with charge $Q_i^a$. In component notation, the
supersymmetry transformations of the vector multiplet are: 
\begin{eqnarray}
\delta v_0^a&=&i \left(\overline{\epsilon}_+ \lambda_+^a 
+\overline{\epsilon}_-\lambda_-^a ~+~\epsilon_+\overline{\lambda}_+^a
~+~\epsilon_-\overline{\lambda}_-^a \right), \nonumber \\
\delta v_1^a&=& i\left( \overline{\epsilon}_+ \lambda_+^a -  
\overline{\epsilon}_- \lambda_-^a+\epsilon_+ \overline{\lambda}_+^a
-\epsilon_- \overline{\lambda}_-^a \right), \nonumber \\
\delta \sigma^a&=&-i \sqrt{2} \overline{\epsilon}_+
\lambda_-^a -i\sqrt{2}\epsilon_-\overline{\lambda}_+^a,\nonumber \\
\delta \overline{\sigma}^a&=&-i \sqrt{2}\epsilon_+\overline{\lambda}_-^a -i
\sqrt{2}\overline{\epsilon}_-\lambda_+^a, \nonumber \\
\delta D^a&=&- \overline{\epsilon}_+(\partial_0 - \partial_1)\lambda_+^a
-\overline
{\epsilon}_-(\partial_0 + \partial_1)\lambda_-^a \\ 
&+&\epsilon_+(\partial_0 -
\partial_1) \overline{\lambda}_+^a +\epsilon_-(\partial_0 + \partial_1)
\overline{\lambda}_-^a, \nonumber \\
\delta \lambda_+^a&=&i\epsilon_+D^a+\sqrt{2}
(\partial_0+\partial_1)\overline{\sigma}^a
\epsilon_- -v^a_{01} \epsilon_+, \nonumber \\
\delta \lambda_-^a&=&i \epsilon_-D^a+\sqrt{2}(\partial_0- \partial_1)\sigma^a
\epsilon_+ +v^a_{01}\epsilon_-, \nonumber \\
\delta \overline{\lambda}_+^a&=&-i \overline{\epsilon}_+D^a
+\sqrt{2}(\partial_0 + \partial_1) \sigma^a\overline{\epsilon}_- 
-v^a_{01}\overline{\epsilon}_+ \nonumber, \\
\delta \overline{\lambda}_-^a&=&-i \overline{\epsilon}_-D^a
+\sqrt{2}(\partial_0 - \partial_1)\overline{\sigma}^a \overline{\epsilon}_+ 
+v_{01}^a\overline{\epsilon}_- \ ,\nonumber
\label{vectortrans}
\end{eqnarray}
where $\epsilon_\pm$ and $\overline{\epsilon}_\pm$ are 
the Grassman parameters for SUSY transformations. 
The transformation rules for the chiral multiplet are:
\begin{eqnarray}
\delta \phi_i&=& \sqrt{2} (\epsilon_+ \psi_{-i}-\epsilon_- \psi_{+i}),
\nonumber \\
\delta \psi_{+i}&=&i \sqrt{2}(D_0 + D_1) \phi_i
\overline{\epsilon}_-+\sqrt{2}\epsilon_+F_i-2Q_i^a \phi_i
\overline{\sigma}^a\overline{\epsilon}_+, \nonumber \\
\delta \psi_{-i}&=&-i \sqrt{2}(D_0 - D_1) \phi_i \overline{\epsilon}_+
+ \sqrt{2} \epsilon_- F_i+2Q_i^a \phi_i \sigma^a\overline{\epsilon}_-,
 \\
\delta F_i&=&-i \sqrt{2} \overline{\epsilon}_+(D_0 - D_1) \psi_{+i}
-i \sqrt{2}\overline{\epsilon}_-(D_0 + D_1)\psi_{-i} \nonumber \\
& & \mbox{} +2Q_i^a(\overline{\epsilon}_+ \overline{\sigma}^a \psi_{-i}
 + \overline{\epsilon}_- \sigma^a\psi_{+i} )
+2iQ_i^a\phi_i(\overline{\epsilon}_- \overline{\lambda}_+^a- 
\overline{\epsilon}_+ \overline{\lambda}_-^a)
\label{chiraltrans}
\end{eqnarray}

The supersymmetric bulk action can be written as a sum of four terms, 
\begin{equation}
S=S_{ch}+S_{gauge}+S_{W}+S_{r,\theta}
\label{bulkaction}
\end{equation}
The terms on the right hand side are, respectively: the kinetic term for
the chiral superfields; the kinetic terms for the vector superfields; the
superpotential interaction; and the Fayet-Iliopoulos and theta terms. 
$S_{ch}$ is: 
\begin{eqnarray}
S_{ch}&=&\sum_i\int d^2x \left\{ 
- D_{\alpha}\overline{\phi}_i D^{\alpha} \phi_i
+i \overline{\psi}_{-i}( \stackrel{\leftrightarrow}{D_0}
+\stackrel{\leftrightarrow}{D_1})\psi_{-i} \right.
+i \overline{\psi}_{+i}
( \stackrel{\leftrightarrow}{D_0}
-\stackrel{\leftrightarrow}{D_1})
\psi_{+i} \nonumber \\
& & \mbox{}+~|F_i|^2 ~-~2\sum_a\overline{\sigma}^a\sigma^a 
(Q_i^a)^2\overline{\phi}_i
\phi_i~-~\sqrt{2}\sum_a Q_i^a(\overline{\sigma}^a\overline{\psi}_{+i}\psi_{-i}~
+~\sigma^a
\overline{\psi}_{-i}\psi_{+i}) \nonumber \\ 
& & \mbox{}~+~D^aQ_i^a\overline{\phi}_i\phi_i 
-i\sqrt{2}\sum_aQ_i^a\overline{\phi}_i(\psi_{-i}
\lambda_+^a~-~\psi_{+i} \lambda_-^a) \nonumber\\
& & \left. \mbox{}~-~i \sqrt{2}Q_i^a\phi_i
(\overline{\lambda}_-^a
\overline{\psi}_{+i}~-~\overline{\lambda}_+^a\overline{\psi}_{-i} ) 
\right\}
\label{kechiral}
\end{eqnarray}
where
\begin{equation}
	A \stackrel{\leftrightarrow}{D_i} B \equiv 
	{1\over2} (A D_i B - (D_iA)B)\ .
\label{symder}
\end{equation}
This symmetrized form of the fermion kinetic term
is Hermitian in the presence of a boundary.
Meanwhile, $S_{gauge}$ is:
\begin{eqnarray}
S_{gauge}&=&\sum_a{1 \over e_a^2}\int d^2x\left\{{1 \over 2}(v_{01}^a)^2 + 
{1 \over 2}(D^a)^2 - \partial_{\alpha}\sigma^a \partial^{\alpha} 
\overline{\sigma}^a \right.  \nonumber \\
&&+  \left. i \overline{\lambda}_+^a
(\stackrel{\leftrightarrow}{\partial_0}-
\stackrel{\leftrightarrow}{\partial_1})
\lambda_+^a +i \overline{\lambda}_-^a
(\stackrel{\leftrightarrow}{\partial_0}+
\stackrel{\leftrightarrow}{\partial_1})
\lambda_-^a \right\}
\label{kevector}
\end{eqnarray}
The superpotential term is:
\begin{equation}
S_W=-\int d^2x \left( F_i {\partial W \over \partial
\phi_i}~+~{\partial^2 W \over \partial \phi_i\partial\phi_j} \psi_{-i}
\psi_{+j}~+~{\overline F}_i
{\partial \overline{W} \over \partial \overline{\phi}_i}~-~{\partial^2
\overline{W}
\over
\partial\overline{\phi}_i\partial\overline{\phi}_j}
\overline{\psi}_{-i}\overline{\psi}_{+j}
\right)\ .
\label{superpot}
\end{equation}
Finally, the Fayet-Iliopoulos D-term and theta term are:
\begin{equation}
S_{r, \theta}=-r_a\int d^2y D^a~+~{\theta_a\over 2 \pi} \int d^2 y
v_{01}^a\ .
\label{fid}
\end{equation}

We wish to describe a theory with an $N=(2,2)$ superconformal fixed point. 
Thus we wish anomaly-free vector and axial $U(1)$ R-symmetries: these may
be constructed if $\sum_i Q_i^a = 0$ \cite{wittenphases}.  These
R-symmetries may also be used to topologically twist the theory, as we
will discuss below. 

Let us review the manifestation of the target space geometry, and of the
phase structure of the moduli space of compactifications, 
in the class of examples 
which we will use for most of this paper, namely hypersurfaces
in weighted projective space with a single K\"ahler modulus.
The spectrum is a single abelian vector multiplet and $5$ chiral
multiplets. The latter consist of $5$ chiral superfields $\Phi_i$ with
positive charge $Q_{i=1\ldots 5} $ and an additional superfield $\Phi_{6} = P$
with charge $Q_{6} = Q_p = - \sum_{i=1}^5 Q_i$. Furthermore, we choose the 
superpotential
\begin{equation}
	W(\Phi,P) = P G(\Phi)
\end{equation}
where $G$ is a quasi-homogenous  transverse polynomial. There
are four such examples:
a degree five hypersurface in $\BP_{1,1,1,1,1}^4=\BP^4$ 
(the quintic); degree six hypersurface in $\BP_{1,1,1,1,2}^4$;
degree eight hypersurface in $\BP_{1,1,1,1,4}^4$ 
and degree ten hypersurface in $\BP_{1,1,1,2,5}^4$. 

We wish to find the moduli space of supersymmetric ground states, which
should flow to the target space of the infrared CFT. We do so by setting
the bosonic potential energy
\begin{equation}
U=\sum_i F_i^2+\frac{1}{2e^2}D^2+2|\sigma|^2\sum_iQ_i^2|\phi_i|^2
\label{bosonicpe}
\end{equation}
to zero.  We substitute the
equations of motion for
the auxiliary fields $D$ and $F_i$:
\begin{eqnarray}
D&=&-e^2\left(\sum_iQ_i|\phi_i|^2-r\right)\nonumber \\
F_i^*&=&\frac{\partial W} {\partial\phi_i}\ ,
\label{dandf}
\end{eqnarray}
to find that:
\begin{equation}
U=|G(\phi_i)|^2+|p|^2\sum_ii\left|
\frac{\partial G}{\partial \phi_i}\right|^2+
\frac{D}{2e^2}+2|\sigma|^2
\left(\sum_iQ_i^2|\phi_i|^2+Q_p^2|p|^2\right)\ .
\label{bosonicpe1}
\end{equation}
where $p$ and $\phi_i$ represent the 
scalar components of $P$ and $\Phi_{i=1\ldots 5}$
respectively.

Let us begin with the case $r\gg 0$. The $D$ term requires that the
$\phi_i$ cannot all simultaneously vanish. Thus $\sigma = 0$;
transversality of $G$ requires that $p=0$. One must also set the
Fayet-Iliopoulous D-term to zero: 
\begin{equation}
\sum_iQ_i|\phi_i|^2=r\ .
\end{equation}
This condition together with dividing out the $U(1)$ gauge symmetry means
that the $\phi^i$ describe the weighted projective space
$\BP_{Q_1,\ldots,Q_5}^{4}$.  Finally, the condition
$G=0$ means that the $\phi$ live on a degree $|Q_p|$ hypersurface in
$\BP_{Q_1,\ldots,Q_5}^{4}$.

For $r \ll 0$, vanishing of the $D$ term requires that $p\neq 0$.
Transversality of $G$ then implies that all $\phi_i=0$.  This theory has a
unique classical vacuum; the massless excitations are governed by a
superpotential with a degenerate critical point, {\it i.e.} it is a
Landau-Ginzburg theory.  The residual gauge invariance (for instance,
$\BZ_5$ in 
the quintic) in fact implies that it is a Landau-Ginzburg orbifold. At $r
\rightarrow - \infty$ this is believed to be the exactly solvable Gepner
model for the quintic \cite{gepner} (see \cite{psarev}\ for a review and
references).  In this way the trajectory $r\gg 0 \rightarrow r\ll 0$
interpolates between geometric and non-geometric compactifications. 

Beginning with the flat metric on ${\bf \BC}^{5}$ and imposing the gauge
invariance and D-term conditions, one can see that $r$ is essentially the
size of the ambient projective space.  In spacetime this K\"ahler
parameter is complexified by an NS-NS two-form potential; in this model
this flows from the theta angle.  We will show this explicitly in the next
section, but we can note for now that the fact that $\theta$ is a periodic
variable reflects the periodicity of the 2-form flux.  Furthermore one can
show that for $\theta\neq 2\pi n$, the GLSM is nonsingular even at $r = 0$
\cite{wittenphases}. 

\subsection{GLSM with boundary}

Supersymmetric D-branes configurations will preserve four of the eight
spacetime supercharges of the compactification. The boundaries of the
string worldsheet must therefore preserve half of the $N=(2,2)$
superconformal symmetry of the closed strings.  We take this to mean that
boundaries in the corresponding GLSM should also break half of the
supersymmetries. 

We will work on the half-plane $(x_0,x_1)$ with $x_1 \geq 0$, and impose
boundary conditions at $x_1 = 0$. As with the conformal sigma models, the
possible boundary conditions fall into two classes \cite{ooy}, ``A-type''
and ``B-type.'' Roughly these correspond to branes wrapped on special
Lagrangian submanifolds and on holomorphic cycles, respectively. In our
case, ``A-type'' boundary conditions correspond to setting $\epsilon_\pm =
\eta \overline{\epsilon}_\mp$, where $\eta = \pm 1$; ``B-type'' conditions
correspond to $\epsilon_\pm = \eta \epsilon_\mp$. 

The variation of the action in the presence of a
boundary will generate boundary terms in addition
to the bulk terms proportional to the equations of motion.
One chooses boundary conditions on the fields such that
these boundary terms vanish. 
In addition, SUSY variations of the fields will
also generate boundary terms;
upon choosing the preserved supersymmetries one requires
that these boundary terms also vanish.  We list these
boundary terms here; in the next section we
will use them to derive boundary conditions.
\begin{enumerate}
\item
The boundary terms in the action
generated by general variations
of the fields are:
\begin{eqnarray}
\delta_{ord} S_{kin}&=&\int dx^0\left\{
 -\left[ (\partial_1 \phi_i + iQ_iv_1 \phi_i) \delta
\bar{\phi}_i+(\partial_1 \bar{\phi}_i - iQ_iv_1 \bar{\phi}_i) \delta \phi_i
\right] \phantom{{1\over2}}\right. \nonumber \\
& & \left.\mbox {} +{i \over 2}\left[(\bar{\psi}_{-i}\delta\psi_{-i}-\psi_{+i} 
\delta \bar{\psi}_{+i})-(\bar{\psi}_{+i}\delta\psi_{+i}-\psi_{-i}\delta
\bar{\psi}_{-i}) \right]\right\}, \nonumber \\
\delta_{ord} S_{gauge}&=&{1 \over e_a^2}\int dx^0 \left\{-v_{01}^a\delta v_0^a
-\left[(\partial_1 
\bar{\sigma}^a)\delta\sigma^a+(\partial_1\sigma^a)\delta \bar{\sigma}^a 
\right]\phantom{{1\over2}}\right.\nonumber \\ 
& & \mbox{} \left.+{i \over 2} \left[ (\bar{\lambda}_-^a\delta
\lambda_-^a -\lambda_+^a\delta\bar{\lambda}_+^a)-(
\bar{\lambda}_+^a\delta\lambda_+^a-\lambda_-^a\delta \bar{\lambda}_-^a
\right)] \right\}\nonumber \\
\delta_{ord} S_{r, \theta}&=&-{\theta_a\over 2\pi}\int dx^0 ~\delta v_0^a
\label{eighteen}
\end{eqnarray}
\item
The boundary terms generated by the transformation
(\ref{vectortrans}),(\ref{chiraltrans}) are:
\begin{eqnarray}
\delta_{susy} S_{kin}&=&\int dx^0 \left\{ {1\over\sqrt{2}} \left[ 
(D_0 \phi_i)(\bar{\epsilon}_+ \bar{\psi}_{-i}+\bar{\epsilon}_-
\bar{\psi}_{+i})-(D_0
\bar{\phi_i})(\epsilon_+ \psi_{-i}+\epsilon_-\psi_{+i})\right]\right.
\nonumber \\
& & \mbox{}+{1 \over \sqrt{2}}\left[(D_1\phi_i)(\bar{\epsilon}_+
\bar{\psi}_{-i}-\bar{\epsilon}_-\bar{\psi}_{+i}) -
(D_1 \bar{\phi_i})(\epsilon_+\psi_{-i}-\epsilon_-
\psi_{+i})\right]\nonumber \\
& & \mbox {} + iQ_i^a\left[ \bar{\phi_i}\sigma^a \epsilon_+ \psi_{+i}
+\phi_i\bar{\sigma}^a\bar{\epsilon}_+\bar{\psi}_{+i}+
\bar{\phi_i} \bar{\sigma}^a\epsilon_-\psi_{-i}
+\phi_i\sigma^a\bar{\epsilon}_-\bar{\psi}_{-i}
\right]
\nonumber \\
& & \mbox{} +\left. {i\over\sqrt{2}}\left[(\bar{\epsilon}_+\psi_{+i}-
\bar{\epsilon}_-\psi_{-i})\bar{F}_i~+~(\epsilon_+ \bar{\psi}_{+i}
-\epsilon_- \bar{\psi}_{-i})F_i\right] \right\}\nonumber \\
\delta_{susy} S_{gauge}&=&{1 \over e_a^2}\int dx^0\left\{
{i\over\sqrt{2}}\left[(\partial_0
\sigma^a)(\epsilon_+ \bar{\lambda}_-^a-\bar{\epsilon}_-
\lambda_+^a)-(\partial_0\bar{\sigma}^a)(\epsilon_-\bar{\lambda}_+^a
-\bar{\epsilon}_+\lambda_-^a)\right]\right.\nonumber \\
& & \mbox{}+{ i \over \sqrt{2}} \left[(\partial_1
\sigma^a)(\epsilon_+\bar{\lambda}_-^a+\bar{\epsilon}_-
\lambda_+^a)+(\partial_1 \bar{\sigma}^a)(\epsilon_-\bar{\lambda}_+^a
+\bar{\epsilon}_+\lambda_-^a)\right]\nonumber \\
& & \mbox{}-{i \over 2}v_{01}^a\left[\epsilon_+\bar{\lambda}_+^a
+\epsilon_-\bar{\lambda}_-^a+\bar{\epsilon}_+\lambda_+^a
+\bar{\epsilon}_-\lambda_-^a\right]\nonumber \\
& & \mbox{} + \left. {D^a\over 2}\left[\epsilon_+\bar{\lambda}_+^a-\epsilon_-
\bar{\lambda}_-^a+\bar{\epsilon}_-\lambda_-^a-\bar{\epsilon}_+
\lambda_+^a\right] \right\}\nonumber \\
\delta_{susy} S_W&=&i\sqrt{2}\int dx^0\left[{\partial W\over\partial\phi_i} 
(\bar{\epsilon}_-\psi_{-i}-\bar{\epsilon}_+\psi_{+i})+{\partial\bar {W}
\over\partial\bar{\phi_i}}(\epsilon_-\bar{\psi}_{-i}-\epsilon_+
\bar{\psi}_{+i})\right] \nonumber \\
\delta_{susy} S_{r, \theta}&=&{-i\theta_a\over 2\pi}\int dx^0
\left[\bar{\epsilon}_+\lambda_+^a+\epsilon_+\bar{\lambda}_+^a
+\bar{\epsilon}_-\lambda_-^a+\epsilon_-\bar{\lambda}_-^a\right]
\label{nineteen}
\end{eqnarray}
\end{enumerate}

\subsection{Landau-Ginzburg theories with boundary}

Finding appropriate boundary conditions is fairly
complicated; in addition, we would like to know
their physical import.  One tool is to try and
understand sensible boundary conditions for
$r \gg 0$ and $r \ll 0$, where 
gauge theory/worldsheet instanton corrections
are small \cite{wittenphases,psarev}.  
The former limit is described by a nonlinear sigma model
and we will discuss this in the next section.
The latter limit is described by a Landau-Ginzburg
theory, and we review here supersymmetric
boundary conditions for such theories 
\cite{stt,boundarylg,HIV}.\footnote{Previous work
on Landau-Ginzburg theories with boundary
can be found in \cite{warner}.}

\subsubsection{A-type boundary conditions}
Consider a Landau-Ginzburg model with $n$ chiral
superfields $\Phi_i$ and arbitrary superpotential $G(\Phi)$. 
For A-type boundary conditions, we impose $n$ independent conditions
\begin{equation}
f_a(\phi,\overline{\phi})=0\quad,
\end{equation}
where $f_a$ are real functions. We will use the indices $i,j,\cdots$ to
denote the superfields and the indices $a,b,c,\cdots$ to indicate the
boundary conditions.
Let $\Sigma$ denote the sub-manifold in 
$\BC^n$ (with complex coordinates
$\phi_i$ and $\overline{\phi}$) obtained by imposing
these conditions. 
We will in addition impose the compatibility condition:
\begin{equation}
\{f_a(\phi,\overline{\phi}),f_b(\phi,\overline{\phi})\}_{PB}=0\ ,
\end{equation}
where:
\begin{equation}
	\{A,B\} = g^{i\bar{j}} \left( \partial_i A 
	\bar{\partial}_{\bar{j}} B -
	\bar{\partial}_{\bar{j}} A
	\partial_i B\right) \ .
\end{equation}
We will assume that on $\Sigma$, the normals $\vec{n}_a
\equiv (\partial_i f_a,\overline{\partial}_{\bar{i}} f_a)$ 
span the normal bundle ${\cal N}\Sigma$. The vanishing of the Poisson
bracket can be rewritten as
\begin{equation}
\vec{n}_a \cdot \vec{t}_b =0
\end{equation}
where 
$\vec{t}_b\equiv 
(\partial_i f_b,-\overline{\partial}_{\bar{i}} f_a)$
are tangent vectors to the curve $f_b=0$. It follows that they span
the tangent bundle $T\Sigma$. $\Sigma$ is thus a {\it Lagrangian
submanifold} of $\BC^n$ by construction 
\cite{harveylawson}. The induced
metric (first fundamental form) on $\Sigma$ is given by
\begin{equation}
h_{ab} = \vec{t}_a \cdot \vec{t}_b = \vec{n}_a \cdot \vec{n}_b \quad.
\end{equation}

The following set of additional
boundary conditions on the fields in the LG model
are consistent with the vanishing of the 
boundary terms which occur in the
general and supersymmetric variations of the LG Lagrangian.
Define
$\chi_{\pm a}\equiv {{\partial f_a}\over{\partial\phi_i}}\ \psi_{\pm i}$.
Then: 
\begin{eqnarray}
\chi_{+a} + \eta \overline{\chi}_{-a} =0\quad, \\
\left(
\left[{{\partial f_a}\over{\partial\phi}_i}
\partial_1\phi_i
- {{\partial f_a}\over{\partial\overline{\phi}_i}}
\partial_1\overline{\phi}_i\right]
-i K_{abc}\ \chi_{-}^b\ \overline{\chi}_{-}^c \right) =0\\
\left\{f_a(\phi,\overline{\phi}),G(\Phi)-\overline{G}(\overline{\phi})
\right\}_{PB}=0
\end{eqnarray}
The complex conjugate conditions are also hold.
$K_{abc}$ is the extrinsic curvature tensor (second fundamental form)
given by
\begin{equation}
K_{abc} =-\left [
{{\partial f_c}\over{\partial\phi_i}}
{{\partial f_b}\over{\partial\phi_j}}
{{\partial^2f_a}\over {\partial\phi_i\partial\phi_j}}
- {{\partial f_c}\over{\partial\overline{\phi}_i}}
 {{\partial f_b}\over{\partial\phi_j}}
{{\partial^2f_a}\over {\partial\phi_i\partial\overline{\phi}_j}}
- {{\partial f_c}\over{\partial\phi_j}}
 {{\partial f_b}\over{\partial\overline{\phi}_i}}
{{\partial^2f_a}\over {\partial\phi_i\partial\overline{\phi}_j}}
+{{\partial f_c}\over{\partial\overline{\phi}_i}}
{{\partial f_b}\over{\partial\overline{\phi}_j}}
{{\partial^2f_a}\over{\partial\overline{\phi}_i\partial\overline{\phi}_j}}
\right]\ .
\end{equation}

This full set of boundary conditions
is equivalent to requiring that $\Sigma$
be Lagrangian.  Without a superpotential, this
corresponds to the microscopic (worldsheet)
realisation of situations considered by
Harvey and Lawson \cite{harveylawson}.
In the presence of a superpotential, there 
is an additional condition that the real
conditions $F_a$ have a vanishing Poisson bracket with $(G-\overline{G})$. 
This suggests that one must choose one of the conditions
to be $F=(G-\overline{G})-i c$ where $c$ is a real constant,
as there can only $n$ independent commuting constants of motion
in a 2n real-dimensional phase space

\subsubsection{B-type boundary conditions}

Under B-type boundary conditions, the unbroken $N=2$ supersymmetry
is given by the condition
\begin{equation}
\epsilon_+ = \eta\ \epsilon_-\quad,
\end{equation}
where $\eta=\pm1$. The following linear boundary conditions
were constructed in the LG model\cite{stt} 
\begin{eqnarray}
(\psi_{+i} + \eta {B_i}^j\psi_{-j})|_{x=0}=0\quad, \nonumber  \\
\partial_1(\phi_i + {B_i}^j\phi_j)|_{x=0}=0\quad, \nonumber  \\
 \partial_0(\phi_i - {B_i}^j\phi_j)|_{x=0}=0\quad, \nonumber  \\
\left.\left({{\partial G}\over{\partial\phi_i}} + {B^*_i}^j {{\partial
G}\over{\partial\phi_j}}\right)\right|_{x=0}=0 \quad,
\end{eqnarray}
where the boundary condition is specified by a hermitian matrix
$B$ which satisfies $B^2=1$.  Since $B$ squares to one, its
eigenvalues are $\pm1$.  An eigenvector of $B$ with eigenvalue of
$+1$ corresponds to a Neumann boundary condition and $-1$
corresponds to a Dirichlet boundary condition. Associated with
every eigenvector with eigenvalue $+1$, there is a non-trivial
condition involving the superpotential which is given by the last
of the above boundary conditions. 

More general possibilities are given by boundary conditions corresponding to
a holomorphic submanifold $\Sigma\subset\BC^n$
defined  by the tranverse intersection of the $r$ conditions
\begin{equation}
f_m(\phi)=0 \quad,\ (m=1,\ldots,r) 
\end{equation}
where $f_m$ are quasi-homogeneous holomorphic functions of the $\phi_i$. 
Under supersymmetric 
variation with $\epsilon^+=\eta \epsilon^-$, one obtains the conditions
\begin{equation}
\sum_i n_m^i(\phi) (\psi_{+i} -\eta \psi_{-i})=0
\end{equation}
where $n_m^i \equiv (\partial f_m/\partial\phi_i)$ are the (holomorphic)
normals to the surface $f_m=0$. 
Let $t_i^a(\phi)$ ($a=1,\ldots,n-r$) be a basis
of tangent vectors to $\Sigma$
such that in local holomorphic coordinates $z_a$, 
$t^a_i ={{\partial\phi_i}\over{\partial z_a}}$.
One needs to impose 
further boundary conditions in order to cancel fermionic
boundary terms arising in the ordinary variation of the action:
\begin{equation}
\eta^{i\bar{j}}t^a_i(\phi) (\overline{\psi}_{+j} +\eta \overline{\psi}_{-j})=0
\end{equation}
Supersymmetric variation of the above equation leads to the conditions
\begin{eqnarray}
\eta^{i\bar{j}} \left(t^a_i(\phi) \partial_1\overline{\phi}_j +{i\over2}
{{\partial t^a_i}\over{\partial\phi_k}} (\psi_{+k} -\eta \psi_{-k})
(\overline{\psi}_{+j}+\eta \overline{\psi}_{-j})\right)=0 \\
t^a_i(\phi) {{\partial G}\over{\partial\phi_i}} =0
\label{supbc}
\end{eqnarray}
The first equation can be  rewritten in the following  form
\begin{equation}
 \left(t^a_i(\phi) \eta^{i\bar{j}}\partial_1\overline{\phi}_j\right)
- \chi^{ab}_m\ \tau_b\ \overline{\nu}^m =0
\end{equation}
where $\chi^{ab}_m$ is the extrinsic curvature of the submanifold
(second fundamental form) given by (see ref. \cite{candelas} for
a discussion)
$$
\chi^{ab}_m \equiv t^a_i t^b_j{{\partial^2 f_m}\over{\partial\phi_i
\partial\phi_j}}
$$
and $h_{a\overline{b}}\equiv \vec{t}_a \cdot \vec{t}^*_b$
is the induced metric (first fundamental form). 
We have also defined fermionic linear combinations $\tau_a$ and $\nu^m$
$$
({\psi}_{+i} -\eta {\psi}_{-i})=t_i^a \tau_a\quad,$$ 
$$(\overline{\psi}_{+i} +\eta \overline{\psi}_{-i})\eta^{\bar{i}j}=
n_m^j\overline{\nu}^m\quad.$$
These are fermionic combinations
which are sections of the tangent bundle and normal bundle
respectively. The boundary terms under ordinary variations of the LG
action vanish for the above choice of boundary conditions.

The boundary conditions involving the superpotential given by eqn.
(\ref{supbc}) is always satisfied if one chooses one of the boundary
conditions to be $G=0$. For instance, all examples of B-type
boundary conditions in LG models considered
in \cite{stt}) can be seen to imply $G=0$. This requirement has also
been observed independently  in \cite{HIV}. This is the analogue
of $(G-\overline{G})=0$ condition seen in A-type boundary conditions.

\subsection{Topological Aspects of the GLSM}

As already mentioned, the GLSM is not conformally invariant and
flows to a conformally invariant theory in the infrared (IR) 
limit\cite{wittenphases}. It follows that one must be careful in
naively extrapolating results in the GLSM to the conformally
invariant fixed point. For example, in the NLSM limit of the GLSM
for the quintic, the metric is given by the pullback of
the Fubini-Study metric on $\BP^4$. This metric is clearly not the 
correct one.  The fact that both the GLSM and its IR fixed
point have worldsheet $(2,2)$ supersymmetry is quite useful
in obtaining some control. To be precise, by appropriately twisting the
theories, one constructs topological theories whose observables are
insensitive to such differences between the two theories and one
can make predictions.

There are two possible twists of the (Euclidean) $(2,2)$ model: the A-twist 
corresponds to the case when the supersymmetry charge: $Q=Q_-+\overline{Q}_+$
(where $Q_\pm=\int G_\pm$ are the supersymmetry charges associated
with the supersymmetry generators $G_\pm$.)
becomes a scalar and the B-twist is where $Q=\overline{Q}_- + \overline{Q}_+$
is a scalar. Physical states of the topological theory correspond
to cohomology classes of $Q$ and the observables are given by
correlation functions of vertex operators that are Q-closed.
Observables of the topological A-model vary holomorphically in
$t=\frac\theta{2\pi} + i r$ while those in the B-model are independent
of $t$. Correlation functions in the A-model can receive corrections
from gauge theory instantons which do not quite coincide with
the worldsheet instantons corrections seen in the conformally
invariant NLSM. The difference arises because the instanton
moduli space for the GLSM  is compact while that of the NLSM is 
non-compact\cite{wittenphases}. 
However, it has been shown that singularity structure of the moduli
space is correctly predicted in the GLSM\cite{wittenphases,morple}.

For the case of GLSM with boundary, one may hope to apply similar
techniques. As has been pointed out earlier\cite{ooy},
A-type boundary conditions
are compatible with the A-twist and B-type boundary conditions with
the B-twist in the open-string channel. 
For instance, in the topological
A-model, the $\sigma$ field of the vector multiplet is Q-closed.
This can be easily seen by the fact that $\delta \sigma=0$ under
the supersymmetry transformation generated by $Q$ i.e.,
$\epsilon_+=\overline{\epsilon}_-$. Further, in the NLSM limit
(c.f. sec. 3) we will see that
$$
\sigma = -{{\sum_i Q_i \overline{\psi}_{+i}\psi_{-i}}\over{\sqrt2
K[\phi]}} 
$$
In the topological A-model, 
$\overline{\psi}_{+i}$ is a $(1,0)$ form 
(to be precise, a section of $\Phi^*(T^{1,0}(X))$)
and $\psi_{-i}$ a $(0,1)$ form and hence $\sigma$ is proportional
to the K\"ahler form $\omega$ on the Calabi-Yau manifold. The proportionality
constant $K[\phi]$ can be seen to be non-vanishing everywhere. It is known
that A-branes wrap special Lagrangian submanifolds of the Calabi-Yau 
manifold. Lagrangian submanifolds satisfy the condition that the restriction
of the K\"ahler form $\omega$ to the submanifold vanishes. Thus, 
for A-type boundary conditions, it follows that
\begin{equation}
\sigma|_{x^1=0} =0\quad.
\end{equation}
We will see that our analysis in the sequel will be consistent 
with this condition. 

Just as in the closed string case, the interpretation of results
in the case of GLSM with boundary should be done with care.
For instance, in the case of A-type branes,
worldsheet instanton effects lead to a stringy
notion of the topology of the cycle
which can differ significantly from the
topology computed by geometric means \cite{kklm2}.
For the case of B-branes, as we move around
in the K\"ahler moduli space, the branes
can undergo monodromy transformations; thus,
if one writes down boundary conditions far from
the large-radius limit and tries to understand
it by following that boundary condition out
to the large-radius limit, the result can depend
on the path one takes. It is also possible that
a boundary state at the Gepner point has no stable large-radius analog
\cite{doug1,dfr}.
Finally, although taking monodromies and
lines of marginal stability into account when studying D-branes far
from the large radius limit is nontrivial, some progress has been made 
\cite{quintic,doug1,dfr} It would be interesting to study
these effects in the GLSM.

\section{The Nonlinear Sigma Model}

Our eventual goal is to use the boundary GLSM as a tool
for understanding the boundary CFT to which it should flow in the infrared.
To begin with, we would like to understand
what a given set of boundary conditions for the GLSM
might correspond to in the infrared.

As discussed in \cite{wittenphases}, the theory
flows to strong coupling in this limit.  It is therefore
tempting to simply take the limit $e^2 \rightarrow \infty$
and use these results as a physical guide.  In particular,
in this limit the gauge kinetic terms drop out; upon
integrating out the nonpropagating gauge fields one
is left with a nonlinear sigma model.

We will now look for consistent
boundary conditions for the one-modulus examples
in the limit $r \gg 0$.  We hope this will provide 
a simple guide to
finding boundary conditions for finite $e^2$,
as well as a crude guide to the infrared physics of the GLSM.
We begin by describing the results of integrating
out the vector multiplets; along the way we will
have to add a contact term to reproduce sensible
$N=2$ NLSM results.  Following this we will discuss
A-type boundary conditions and Neumann B-type
boundary conditions in this limit.

\subsection{$e^2 \rightarrow \infty$ Limit of the Bulk Linear Sigma Model}

In the $e\rightarrow\infty$ limit of the GLSM,
the kinetic energy terms for the vector
multiplet vanish, so that the component fields behave as Lagrange multipliers.
This leads to the following constraints for general $U(1)$ charge.
\begin{enumerate}
\item The D-term constraint:
\begin{equation}
\sum_i (Q_i |\phi_i|^2 - Q_p |p|^2 - r )=0\ ,
\label{condition}
\end{equation}
When $r\gg 0$, $|p|$ is very massive due to 
eq. (\ref{bosonicpe1});
we will set $p = 0$ for the remainder of the section.
\item The constraints imposed by integrating out the
gauginos are:
\begin{equation}
\sum_i Q_i\overline{\phi}_i \psi_{\pm i} =0
\label{lambconstr}
\end{equation}
\item The equations of motion for $\sigma$ and 
$\overline{\sigma}$:
\begin{eqnarray}
\sigma = -{{\sum_i Q_i \overline{\psi}_{+i}\psi_{-i}}\over{\sqrt2
K[\phi]}} 
\label{sigconstr} \\
\overline{\sigma} = -{{\sum_i Q_i \overline{\psi}_{-i}\psi_{+i}}\over
{\sqrt2 K[\phi]}} \label{sigbconstr}\ ,
\end{eqnarray}
where $K[\phi]\equiv\sum_j Q_j^2 |\phi_j|^2$.
\item The equations of motion for the gauge fields:
\begin{eqnarray}
2 K[\phi]\ v_0 = \sum_i Q_i \left[ 
i (\overline{\phi}_i\partial_0 \phi_i - \phi_i \partial_0
\overline{\phi}_i) +
\overline{\psi}_{-i} \psi_{-i} + \overline{\psi}_{+i} \psi_{+i} \right]
\label{gauss}
\\
2 K[\phi]\ v_1 = \sum_i Q_i \left[ 
i (\overline{\phi}_i\partial_1 \phi_i - \phi_i \partial_1
\overline{\phi}_i) -
\overline{\psi}_{-i} \psi_{-i} + \overline{\psi}_{+i} \psi_{+i} \right]
\label{gaussb}
\end{eqnarray}
The equation for $v_0$ is simply Gauss' law.
\item Further supersymmetric variation of the above equations
leads to 
\begin{eqnarray}
-\sqrt2 K[\phi] \lambda_+ = \sum_i Q_i \left[ \overline{\psi}_{-i} 
	(D_0 +D_1) \phi_i +i \overline{F}_i \psi_{+i}\right] 
+ {i\sigma\over{K[\phi]}} 
\sum_i Q_i^2 \phi_i \overline{\psi}_{+i}\nonumber \\
-\sqrt2 K[\phi] \lambda_- = \sum_i 
	Q_i \left[ - \overline{\psi}_{+i} (D_0
-D_1) \phi_i +i \overline{F}_i \psi_{-i}\right] 
+ {i\sigma\over{K[\phi]}} 
\sum_i Q_i^2 \phi_i \overline{\psi}_{-i} 
\label{lamminsol}
\end{eqnarray}
Of course we have integrated out $\lambda$, but these
last equations can be used to take the $e^2 \rightarrow\infty$
limit of equations which are functions of $\lambda$.
\end{enumerate}

In $N=2$ supersymmetric 
type II string compactifications on Calabi-Yau
threefolds, the K\"ahler parameters
are complexified by the fluxes of closed
NS-NS two-form gauge fields $B_{ij}$. In the GLSM, this is
reflected by the presence of a $\theta$-term for every 
Fayet-Iliopoulos D-term \cite{wittenphases}.
This can be seen easily in the
$e^2 \rightarrow \infty$ limit.  Substituting
eqs. (\ref{gauss}),(\ref{gaussb}) into the theta term
$S_\theta$ of eq. (\ref{fid}), we find: 
\begin{eqnarray}
S_{\theta}&=&{\theta\over {2 \pi }} \int d^2 x  \sum_i Q_i\left\{ 
i{{D^B_1 \phi_i D^B_0 \overline{\phi}_i -
D^B_0 \phi_i D^B_1 \overline{\phi}_i}\over{K[\phi]}}\right. \nonumber \\
&& \left.-(\partial_0+\partial_1) 
\left({{\overline{\psi}_{-i} \psi_{-i}}\over
{2K[\phi]}}\right)
 +(\partial_0-\partial_1) \left({{\overline{\psi}_{+i} 
\psi_{+i}}\over{2K[\phi]}} \right)\right\}
\label{Bterm}
\end{eqnarray}
We define $v_0^B$ to the fermion-independent part
of $v_0$ in (\ref{gauss});
$D_0^B$ is the covariant derivative with $v_0$ replaced by
$v_0^B$.

In the case of the quintic, $K=r$ and the bosonic part of
(\ref{Bterm}) corresponds to:
\begin{equation}
	B_{i\bar{j}} = \frac{i\theta}{4\pi r} \eta_{i\bar{j}}\ .
\label{quinticbfield}
\end{equation}
This is closed and topologically nontrivial in $\BP^4$;
it should thus correspond to the NS-NS B-field modulus
for type II compactification on the quintic.
It remains to understand the fermion bilinear term.
Clearly it is a boundary term, and
it can be discarded in the closed-string case.
We claim that in the case at hand it is sensible
to subtract this off, by adding an explicit boundary term:
\begin{equation}
S^{\rm quintic}_{\rm boundary}=\int dx^0\frac{\theta}{4\pi r}
\sum_i\left(\overline{\psi}_{+i}\psi_{+i}+
\overline{\psi}_{-i} \psi_{-i}\right)
\label{bdyxtra}
\end{equation}

We will now consider the case of more general cases involving weighted
projective spaces.
In $d=2,N=2$ supersymmetric NLSMs,
the fermion bilinear terms which scale with $B$
are\cite{zumino}:
\begin{equation}
	\int d^2 x \overline{\psi}^{\bar{j}}\psi^i
	\left(\partial \phi^k H_{\bar{j} i k} + 
	\partial \overline{\phi}^{\bar{k}}
		H_{\bar{j} i \bar{k}} \right)\ ,
\label{hcoupling}
\end{equation} 
where $H = dB$.
In $N=2$ supersymmetric compactifications of type
II string theory on Calabi-Yau threefolds, 
we must have $H = dB = 0$ and so these
bilinear terms must vanish.
Clearly $H=0$ in eq. (\ref{quinticbfield}).
In fact this is true for the B-field
arising from more general $Q_i$.
The general expression for the B-field in the NLSM limit is given by
\begin{eqnarray}
B_{ij} &=& \sum_a {{i\theta_a}\over{2\pi} }
\left({{[(Q^a_i)^2 Q^a_j - Q^a_i (Q^a_j)^2)] \phi_i \phi_j}\over{
4(K^a[\phi])^2 }} \right) \quad,\\
B_{i\overline{j}} &=& \sum_a {{i\theta_a}\over{2\pi} }
\left( {{Q^a_i \eta_{i\overline{j}}}\over { 2K^a[\phi] }} - 
{{[(Q^a_i)^2 Q^a_j + Q^a_i (Q^a_j)^2] \phi_i \overline{\phi}_j}\over
{4(K^a[\phi])^2 }} \right) \quad,
\end{eqnarray}
where $K^a[\phi]\equiv \sum_i (Q_i^a)^2 |\phi_i|^2$.
An explicit calculation of $H_{ij\overline{k}}$ and
$H_{i\overline{j}\overline{k}}$ leads to the vanishing of
the fermion bilinear in eqn. (\ref{hcoupling}) on imposing
the NLSM constraints given by (\ref{condition}) and (\ref{lambconstr}).
This implies that $H$ vanishes on the subspace given by the D-term
constraint.

For the general case of weighted projective spaces, the B-field corresponds 
to a non-constant B-field. Further, $B_{ij}$ and $B_{\bar{i}\bar{j}}$ are
non-vanishing\footnote{One reason to require these to vanish is to
observe that for D-branes, gauge invariance dictates that the
field strength $F$ and the B-field occur (loosely) in the combination
$(F-B)$. For holomorphic connections on B-branes, one has
$F_{ij}=F_{\bar{i}\bar{j}}=0$. Even in the closed string case, it
is preferable to work in this gauge since the complex K\"ahler moduli
involve only $B_{i\bar{j}}$.}.
Is it possible to find a spacetime gauge tranformation
of the form
\begin{equation}
\delta B_{\mu\nu} = \partial_\mu \Lambda_\nu - \partial_\nu \Lambda_\mu
\end{equation}
such that $B_{ij}$ and $B_{\bar{i}\bar{j}}$ vanish? Under such a
gauge transformation, the worldsheet Lagrangian transforms as
\begin{equation}
\delta S_{\rm B-field} = 2 \int d^2x \left\{ \partial_0 (\Lambda_\mu
\partial_1 \phi^\mu) - \partial_1 (\Lambda_\mu 
\partial_0 \phi^\mu) \right\}\quad.
\label{totalder}
\end{equation}
The following choice of gauge transformation does this:
\begin{eqnarray}
\Lambda_i &=& \sum_a \left({{i\theta_a Q^a_i \phi_i}\over{8\pi K^a[\phi]}} 
- {{i\theta_a \phi_i}\over{8\pi r_a}}\right) \\
\Lambda_{\bar i} &=& \sum_a 
\left(-{{i\theta_a Q^a_i \overline{\phi}_i}\over{8\pi K^a[\phi]}}
+ {{i\theta_a \overline{\phi}_i}\over{8\pi r_a}}\right) 
\end{eqnarray}
Of course, this also modifies $B_{i\overline{j}}$ to the following form
\begin{equation}
B_{i\overline{j}} =
\sum_a \left(\frac{i\theta_a}{4\pi r_a} \right)\eta_{i\bar{j}}\ ,
\end{equation}
which is a constant B-field as in the quintic! There are total
derivative pieces given by the gauge tranformation (see eqn.
(\ref{totalder}). We discard
them and this corresponds to a further addition to the existing contact term 
we obtained by discarding the fermion bilinears. 
For the one-modulus case, the final contact term 
takes the following simple form 
\begin{equation}
S^{NLSM}_{\rm boundary}= {{i\theta}\over{4\pi r}} \int dx^0\ 
\sum_i (\phi_i D_0\overline{\phi}_i - \overline{\phi}_iD_0 \phi_i)
\label{nlsmcontact}
\end{equation}
where we have used eqn. (\ref{gauss}) to rewrite the fermion bilinear in
terms of the bosonic fields.

The attentive reader may observe that the requirement that
$B_{ij}$ and $B_{\bar{i}\bar{j}}$ vanish does not completely
fix the required gauge transformation $\Lambda_\mu$. An additional
requirement is that the boundary conditions in the NLSM and the LSM
agree. Our choice of gauge
transformation given above precisely achieves this.
The important point to note is that the choice of boundary conditions in
the NLSM is dictated by the form of the B-field, whereas in the LSM it is the
contact term (i.e., the terms we discarded in the NLSM limit as total
derivatives) which dictates the choice of boundary conditions. 
This
will become clearer in the following sections. 

As a final note, such a term arises
in an almost identical context in \cite{gut}.
In that work the issue is the construction of
a BRST-invariant vertex operator for the NS-NS B-field
on the disc, for D-instantons in flat space.
Whether the vertex operator describes a fluctuation
$\delta B$ for which $\delta H = 0$, or one for
which $\delta H$ is nonzero, one must add
a ``contact term'' on the boundary which is a fermion
bilinear, in order that the integrated
operator is BRST invariant.  If one writes this
boundary term as a total derivative in the
interior of the disc,
the total fermion part of the vertex operator
has the right symmetry structure in its Lorentz indices to
describe a fluctuation of $H$.

\subsection{A-type boundary conditions in the NLSM}

In Sec. 2, we considered a LG model with $n$ chiral
superfields and arbitrary superpotential. The same techniques 
give us sensible A-type boundary conditions for the $n$-chiral
superfields in the $e^2 \rightarrow \infty$ NLSM.
In this limit, the following boundary conditions
lead to the vanishing of (\ref{eighteen}) and (\ref{nineteen}),
in the limit of infinite gauge coupling:
\begin{eqnarray}
F_a(\phi,\overline{\phi})&=&0\quad, \\
\chi_{+a} + \eta \overline{\chi}_{-a} &=&0\quad, \\
\left(
\left[{{\partial F_a}\over{\partial\phi}_i}
D_1\phi_i
- {{\partial F_a}\over{\partial\overline{\phi}_i}}
D_1\overline{\phi}_i\right]
-i K_{abc}\ \chi_{-}^b\ \overline{\chi}_{-}^c \right) &=&0\\
\left\{F_a(\phi,\overline{\phi}),W(\Phi)-\overline{W}(\overline{\phi})
\right\}_{PB}&=&0
\end{eqnarray}

We may use the equations of motion for the vector multiplet to
infer sensible boundary condiitions on the
component fields, via eqs. (\ref{sigconstr})-(\ref{lamminsol}).
For $\sigma$ and $\lambda$,
\begin{eqnarray}
\sigma|_{x^1=0}=\overline{\sigma}|_{x^1=0}&=&0 \\
(\lambda_+ + \eta \overline{\lambda}_-)|_{x^1=0} &=&0
\label{nlsmgauginobound}
\end{eqnarray}
This implies the following boundary condition on the twisted chiral
superfield (with $\theta^+=\eta \overline{\theta}^-$)
\begin{equation}
\Sigma|_{x^1=0}=0
\end{equation}
Indeed, it happens that $v_0=0$ in the NLSM limit.  This
is consistent with supersymmetry since
$\delta_{susy}v_0=0$ using (\ref{nlsmgauginobound}).
Thus, one can choose the gauge condition 
\begin{equation}
\delta v_0 =0
\end{equation}
in the LSM.

\subsection{B-type boundary conditions in the NLSM}

Recall that B-type boundary conditions are defined by requiring that 
$\epsilon_+ = \eta \epsilon_-$.
In this case the $\theta$ term complicates the story.
We will begin with the case $\theta = 0$.

\subsubsection{$\theta=0$}

As we stated above, we will work with fully
Neumann boundary conditions in this section.
Let us begin with the boundary condition on $\psi$
(justified {\it a posteriori}):
\begin{equation}
(\psi_+ +\eta \psi_-)|_{x^1=0} = 0\ .
\end{equation}
The supersymmetric variations of this condition lead to:
\begin{eqnarray}
\left.\left(D_1\phi-i\eta({{\sigma-\overline{\sigma}}
	\over{\sqrt2}})Q\phi\right)
\right|_{x^1=0}&=&0 \\
F|_{x^1=0}=\left.{{\partial \overline{W}}\over{\partial\overline{\phi}}}
\right|_{x^1=0}&=&0 \ .
\end{eqnarray}
The vanishing of boundary terms in the ordinary variations of the action
requires further that:
\begin{eqnarray}
D_1\phi|_{x^1=0}=0 \\
(\sigma -\overline{\sigma})|_{x^1=0}=0\ .
\label{btypesigma}
\end{eqnarray}
It is easy to see that the conditions on $\sigma$ are
consistent with eqs. (\ref{sigconstr}),(\ref{sigbconstr}).

The remaining boundary conditions on the vector
multiplet can be derived either from
the solutions in eqs. (\ref{sigconstr})-(\ref{lamminsol}),
or from SUSY variations of (\ref{btypesigma}).
The results are:
\begin{eqnarray}
(\lambda_+ -\eta \lambda_-)|_{x^1=0} &=& 0 \\
\partial_1 (\sigma +\overline{\sigma})|_{x^1=0}&=&0\\
v_{01}|_{x^1=0}&=&0
\end{eqnarray}
Following the notation in \cite{wittenphases},
these conditions can be written in superfield notation as:
\begin{equation}
(\Sigma - \overline{\Sigma})|_{x^1=0}=0\ ,
\end{equation}
where we recall that $\theta^+=\eta\theta^-$ 
on the boundary.  ($\theta$'s are
the Grassmann parameters in $d=2$, $N=2$ superspace.) 
Note that holomorphic boundary conditions on the
complex scalars of the chiral superfields leads to reality
conditions on the scalars in the twisted chiral multiplet;
this might have been
anticipated from mirror symmetry, which exchanges A-type and
B-type boundary conditions, as well as (roughly)
chiral and twisted chiral superfields.

\subsubsection{$\theta\neq 0$}

In the presence of the theta term, $D_1 \phi_i|_{x_1=0} =0$
is clearly no longer the correct bosonic boundary 
condition\footnote{Some of the material in this subsection was 
developed in collaboration with Albion Lawrence.}.  This
is to be expected; even for constant NS-NS B-fluxes
through D-branes in flat space, the boundary conditions for the
string worldsheet are roughly:
\begin{equation}
	\eta_{\mu\nu}\partial_1 X^\nu + B_{\mu\nu}
		\partial_0 X^\nu\ =0\quad.
\label{roughbfieldbc}
\end{equation}
We find the same effect here in the $e^2 \to \infty$ limit.

Ordinary variation of eqn. (\ref{Bterm}) then implies that:
\begin{equation}
\delta S^{ord}_{r,\theta}=-\frac{i\theta}{2\pi r}\int dx^0
\sum_i\left(\partial_0 \phi_i\delta{\bar \phi_i}-
\partial_0
{\bar \phi_i}\delta\phi_i\right)\ ,
\label{vardel}
\end{equation}
where we have ignored the total derivative pieces 
since they have been cancelled by the addition of the contact
term. 
Adding (\ref{vardel}) to the boundary term coming from the ordinary 
variation of the bosonic kinetic energy term in the bulk action, 
we get the new boundary term,
\begin{equation}
\delta S^{ord}_{kin}=\int dy\sum_i\left[\left(D_1\phi_i
+\frac{i\theta}{2\pi r}\partial_0
\phi_i\right)\delta{\bar \phi_i}+c.c\right]
\label{boseord}
\end{equation}
However before we claim that the term in brackets multiplying
$\delta\phi_i$ provides the boundary conditions for the $\phi$'s, 
we note that the boundary conditions may be written in different
ways upto the addition of terms that vanish when we use the fact that
$\partial_0\left(
\sum_iQ_i{\bar \phi_i}\phi_i\right)=\partial_1\left(
\sum_iQ_i{\bar \phi_i}\phi_i\right)=0$, since $\sum_iQ_i{\bar \phi_i}
\phi_i=r$ in the bulk.
For example, eqn. (\ref{boseord}) can be rewritten as
\begin{equation}
\delta S^{ord}_{kin}=\int dy\sum_i\left[\left(D_1\phi_i
+\frac{i\theta}{2\pi r}D^B_0
\phi_i\right)\delta{\bar \phi_i}+c.c\right]\quad,
\label{boseord1}
\end{equation}
where $D_0^B\phi = (\partial_0 +i v_0^B) \phi$.
Care should be taken when we write the boundary conditions
in the full GLSM where $D\neq0$.
The consistent way to do this is to choose boundary
conditions that are closed under supersymmetry. 
Let us begin with the fermionic boundary conditions that are consistent with
with the presence of a constant B-field. 
In such situations fermions obey the rotated boundary condition
\begin{equation}
(\psi_{+i}+\eta e^{i\gamma}\psi_{-i})|_{x^1=0}=0
\label{fermbc}
\end{equation}
where $e^{i\gamma}$ is to be determined later by requiring consistency
with the boundary conditions on the bosons.
Supersymmetric variation of (\ref{fermbc}) implies 
\begin{eqnarray}
D_1\phi_i+\frac{1-e^{i\gamma}}{1+e^{i\gamma}}D_0\phi_i&+&\eta i{\sqrt 2}
Q_i\frac{{\bar \sigma}-e^{i\gamma}\sigma}{1+e^{i\gamma}}\phi_i=0
\label{bosebc}\\
F_i&=&0
\label{supervary}
\end{eqnarray}
where we note that the covariant derivatives involve both the bosonic and
fermionic components of $v_0$ and $v_1$. 
The vanishing of the bosonic boundary variation term eqn. (\ref{boseord})
clearly requires
\begin{equation}
\frac{1-e^{i\gamma}}{1+e^{i\gamma}}=\frac{i\theta}{2\pi r}
\end{equation}
Now by judicious use of the fact that the variation of $D$ is zero and
the explicit expressions for the gauge fields and the $\sigma$ field in
the NLSM limit the boundary condition (\ref{bosebc}) indeed reduces  
to 
\begin{equation}
\left(D_1^B\phi_i+
\frac{i\theta}{2\pi r}D_0^B \phi_i\right)|_{x^1=0}=0
\label{finalcondition}
\end{equation}
This boundary condition is indeed the one suggested by the boundary
variation terms in eqn (\ref{boseord1}) ( with the above fermion
boundary conditions $v_1^F=0$). 
In the GLSM it is eqn. (\ref{bosebc}) which will be the natural
boundary condition on the bosonic boundary fields.  
Eqn. (\ref{finalcondition}) may be obtained by showing that 
\begin{equation}
\left[\frac{1-e^{i\gamma}}{1+e^{i\gamma}}iv_0^F
+i{\sqrt 2}\eta
\frac{{\bar \sigma}-e^{i\gamma}\sigma}{1+e^{i\gamma}}
\right]|_{x^1=0}=0
\label{other}
\end{equation} 
where $v_0^F$ is the fermion bilinear part of $v_0$. 
With these boundary conditions, in the $e^2 \rightarrow\infty$ limit,
all of the boundary terms arising in the 
equations of motion and the supersymmetric variations of the action
cancel out. 

Finally we wish to use the equations of motion
for the vector multiplet to arrive at sensible
boundary conditions for the component fields.
Eq. (\ref{sigconstr}) and (\ref{sigbconstr}) imply the boundary condition:
\begin{equation}
({\bar \sigma}-e^{2i\gamma}\sigma)|_{x^1=0}=0
\label{sigbc}
\end{equation}
This is consistent with eqn. (\ref{other}) and the boundary conditions
for the chiral multiplets.
The SUSY variation of (\ref{sigbc}) requires 
\begin{equation}
(\lambda_+-\eta e^{2i\gamma}\lambda_-)|_{x^1=0}=0\ ,
\label{lambdabc}
\end{equation}
which in turn requires
\begin{equation}
\left.\left\{(1-e^{2i\gamma})iD-(1+e^{2i\gamma})v_{01}
+{\sqrt 2}\eta \partial_1({\bar \sigma}+e^{2i\gamma}\sigma)
\right\}\right|_{x^1=0}=0 \quad.
\end{equation}

Our analysis has been for the case of (all Neumann
boundary conditions) in one-modulus examples
in weighted projective spaces. However, the generalisation to many moduli
case is straightforward and we will not enter into the details here.
We now turn to the case of ``mixed'' boundary conditions, i.e., we
impose Dirichlet boundary conditions on some of the fields.

\subsubsection{``Mixed'' boundary conditions}

We illustrate the general situation by imposing  Dirichlet boundary
conditions on one of the fields, say 
$$
\phi_1=0\quad.
$$ 
Supersymmetric completion requires the condition 
$$
(\psi_{+1} - \eta \psi_{-1})=0\quad.
$$
All other fields have Neumann boundary conditions imposed on them as
in sec. 3.3.2. One can check that all boundary terms in the 
ordinary and supersymmetric variations of the action vanish
as they did in the all Neumann case.

What about the boundary conditions implied on the fields in the vector
multiplet? By considering the expressions for $\sigma$ and
$\overline{\sigma}$ as given in eqns. (\ref{sigconstr}) and
(\ref{sigbconstr}) respectively, it appears that one  cannot
obtain simple boundary conditions as in (\ref{sigbc}) for the all
Neumann case. However, we claim that the problem may be resolved
by requiring that on the boundary the bulk expressions for the
fields in the vector multiplet have to be {\it modified}.
The modification requires that the fields in the vector multiplet
depend only on fields in the chiral multiplet which
have Neumann boundary conditions and not on those with Dirichlet
boundary conditions($\phi_1$, $\psi_{\pm1}$ in our case). 
Note that this is trivially true for the
bosonic part and refers only to the fermionic part such as the bilinear
expression for $\sigma$ in the NLSM limit (see eqn. 
(\ref{sigconstr})). Another way to state this modification is to impose
a modified Gauss law constraint on the boundary such that
\begin{eqnarray}
J_0 &\equiv& 
\sum_i Q_i\left[
i (\overline{\phi}_i D_0 \phi_i - \phi_i D_0
\overline{\phi}_i) +
\overline{\psi}_{-i} \psi_{-i} + \overline{\psi}_{+i} \psi_{+i} \right]
\nonumber \\
&=&Q_1(\overline{\psi}_{-1} \psi_{-1} + \overline{\psi}_{+1} \psi_{+1})
\end{eqnarray}
For the general case, the RHS of the Gauss law constraint  is given by
a summation over fermions in the Dirichlet directions.
Consistency with supersymmetry forces $\sigma$ as
well as $\lambda_\pm$ to also depend only on fields with Neumann
boundary conditions.

Thus, the boundary conditions on the fields in the vector multiplet 
are identical to the all Neumann case considered in section 3.3.2 above.
This will enable us to now carry over these boundary conditions to the
GLSM in a straightforward fashion.

\section{Boundary conditions in the GLSM}

In this section, we propose to derive supersymmetry preserving
boundary conditions for the gauged linear sigma model, that will
define appropriate D-branes wrapping supersymmetric cycles of
the Calabi-Yau. These boundary conditions that we derive for the GLSM, 
would have to be consistent with those in the infra-red limit, i.e 
should appropriately reduce to the ones we have described in the last
section. 

Let us begin by recalling that whereas in the 
non-linear sigma model limit of the GLSM 
the fields in the gauge multiplet become very massive and
effectively decouple from the theory, this is not the case
with the GLSM. As a consequence the analogue of the 
Dirichlet and Neumann boundary conditions in the GLSM are more 
general boundary conditions that depend on fields of the gauge multiplet 
as well. In particular, it is clear that the boundary conditions must
be gauge covariant. 

\subsection{A-type boundary conditions in the GLSM}

In order to extend the boundary conditions from the NLSM
to the GLSM, we
will have to include boundary conditions for the p-field in addition
to the ones we obtained for the
fields in the vector multiplet. One can check that the following  are a
consistent set of boundary conditions in the GLSM at finite gauge
coupling
\begin{eqnarray}
{\rm Im}\ p|_{x^1=0} &=& 0\quad, \\
(\psi_{+p} + \eta \overline{\psi}_{-p})|_{x^1=0}&=&0\quad, \\
{\rm Re}\ D_1  p|_{x^1=0} &=& 0
\end{eqnarray}
All other boundary conditions are as given in the NLSM discussed
earlier. It is interesting that the A-type boundary conditions are
identical in form in both the GLSM in general (taking into account of
course the need to include boundary conditions on the p-field)
and in the LG and CY phases. This is however not the case with the B-type
as is clear even from the considerations of such
boundary conditions in the NLSM limit. 

\subsection{B-type boundary conditions in the GLSM}

\subsubsection{$\theta=0$}

In the NLSM limit, we have seen that for both Dirichlet as well as Neumann
boundary conditions imposed on the matter fields, the boundary
conditions implied on the fields in the vector multiplet are summarised
by the simple boundary condition on the twisted chiral superfield
\begin{equation}
(\Sigma - \overline{\Sigma})|_{x^1=0} =0\quad,
\end{equation}
where we impose $\theta^+=\eta\theta^-$ and
$\overline{\theta}^+=\eta\overline{\theta}^-$ as well in the above condition.
We will continue to require this set of boundary conditions in the
GLSM. However we will also need to impose boundary  conditions on
the $p$-field as well as its supersymmetric partners $\psi_{\pm p}$
such that boundary terms in the ordinary as well as supersymmetric
variation of the GLSM action vanish. 
It is useful to observe at this point that once we
have fixed the above boundary conditions on the fields in the vector
multiplet, the choices of consistent boundary conditions is in fact
identical to that of an extended LG model involving the $p$-field
and the fields $\phi_i$ with superpotential $W=PG(\phi)$
as in section 2 (with the condition that ordinary
derivatives in the LG model  are replaced by covariant derivatives
in the GLSM).  This leads to two possible classes of boundary conditions 
\begin{enumerate}
\item Dirichlet boundary condition on $p$ with $p=0$. Since the
superpotential in the GLSM is given by $W=PG(\phi)$, for this choice,
$F_i^*=(\partial W/\partial \phi_i)= p (\partial G/\partial \phi_i)=0$.
Thus, the condition $F_i=0$ which occurs whenever we impose Neumann
boundary conditions on the $\phi_i$ 
is trivially satisfied. Thus, any boundary
condition (Neumann or Dirichlet) involving 
scalar fields other than $p$ goes through subject to
the condition that all Dirichlet boundary conditions
are specified by homogeneous polynomials.
This includes the all Neumann case which did not
appear in the LG phase. In the LG phase,
(which occurs for large negative $r$) where $p$ has non-vanishing vev,
the boundary condition $p=0$ cannot be imposed. 
This also suggests that this boundary condition is acceptable 
only in the  CY phase where $p=0$ is the ground state condition.
\item Another possibility is that one imposes Neumann boundary
condition on the $p$-field. 
However, the $F_p=0$ condition now requires $G=0$.
Further, one is not allowed to choose to impose Neumann boundary
conditions on individual fields for the quintic at the Fermat point.
A possible consistent choice of boundary conditions at the Fermat point
of the quintic is given by $\phi_1 + \phi_2=0$
and $\phi_i=0$ ($i=3,4,5$). 
\end{enumerate}
One can verify that for both classes of boundary conditions discussed
above, the boundary terms in the ordinary as well as supersymmetric 
variations vanish.

\subsubsection{$\theta\neq 0$}

The strategy that we will pursue is to extend the boundary conditions
we obtained in the NLSM to that of the GLSM. The ordinary and
supersymmetric variations of the kinetic energy terms of the fields in
the vector multiplet will now have to be considered. We will first
consider the contact term which we {\it derived} in the NLSM limit
as given by eqn. (\ref{nlsmcontact}).
\begin{equation}
S^{NLSM}_{\rm boundary}= {{i\theta}\over{4\pi r}} \int dx^0\ 
\sum_i (\phi_i D_0\overline{\phi}_i - \overline{\phi}_iD_0 \phi_i)
\end{equation}
However, before we choose this to be the contact term in the GLSM we
must remember that there may be a need for other terms which vanish in the
NLSM limit. In fact, we do need such a term in order to ensure that
there is a smooth NLSM limit to the boundary conditions chosen in
the GLSM. The full boundary term that  we need in the GLSM turns out be
\begin{equation}
S^{GLSM}_{\rm boundary}= \int dx^0\left\{ {{i\theta}\over{4\pi r}} 
\sum_i (\phi_i D_0\overline{\phi}_i - \overline{\phi}_iD_0 \phi_i)
+ \eta{{\theta}\over{\sqrt{2}\pi r}} {D\over{e^2}}\sigma e^{i\gamma}
\right\}\label{boundterm},
\end{equation}
where $\tan (\gamma/2)=-\theta/2\pi r$.

In the presence of this boundary term, we now list the boundary conditions 
required(using the NLSM as a guide) for cancelling 
ordinary as well as supersymmetric variations of the action. In the
matter sector, these are
\begin{eqnarray}
(\psi_{+i}+\eta e^{i\gamma}\psi_{-i})|_{x^1=0}&=&0\\
\left.\left\{D_1\phi_i+\frac{1-e^{i\gamma}}{1+e^{i\gamma}}
D_0\phi_i+i{\sqrt 2}\eta
Q_i \frac{{\bar \sigma}-e^{i\gamma}\sigma}{1+e^{i\gamma}}\phi_i
\right\}\right|_{x^1=0}&=&0 \label{modbose}\\
p|_{x^1=0}&=&0 \\
(\psi_{+p}-\eta \psi_{-p})|_{x^1=0}&=&0
\end{eqnarray}
where we have included the $p$-field as well as its fermionic partners
$\psi_{\pm p}$.

The boundary conditions on the fields in the vector multiplet in the GLSM
are chosen to be 
\begin{eqnarray}
({\bar\sigma}-e^{2i\gamma}\sigma)|_{x^1=0}=0\\
({\bar\lambda_{+}}-\eta e^{2i\gamma}\lambda_{-})|_{x^1=0}=0
\label{psisiglamg}
\end{eqnarray}
where, in keeping with the results obtained in the NLSM limit,
we have chosen the phase to be twice  the phase in the matter boundary
conditions. The remaining boundary conditions are
\begin{eqnarray}
\left.\left({{v_{01}}\over{D}}+
{\theta\over{2\pi r}}\right)\right|_{x^1=0}&=&0\\
\left.\left(\partial_1(\overline{\sigma} + e^{2i\gamma} \sigma)
-\eta {{\theta}\over{\sqrt{2}\pi r}} 
e^{i\gamma} D\right)\right|_{x^1=0} &=&0
\end{eqnarray}
Notice that the last two boundary conditions are indeed
a convenient split of a single equation (given below) arising from the 
supervariation of eqn. (\ref{psisiglamg}) in order to make boundary terms 
in the ordinary variation vanish. 
\begin{equation}
\left.\left\{(1-e^{2i\gamma})iD-(1+e^{2i\gamma})v_{01}
+{\sqrt 2}\eta \partial_1({\bar \sigma}+e^{2i\gamma}\sigma)
\right\}\right|_{x^1=0}=0
\end{equation}
Hence, they are really dictated by our
insistence that the rotated boundary conditions on $\sigma$ and
$\lambda$ have a {\it consistent} NLSM limit.
The full set of boundary conditions on the fields in the vector
implies the following boundary condition on the twisted chiral
superfield (subject to $\theta^+ = \eta \theta^-$ and
$\overline{\theta}^+ = \eta \overline{\theta}^-$)
\begin{equation}
(\Sigma - e^{-2i\gamma} \overline{\Sigma})|_{x^1=0}=0\quad,
\label{twogamma}
\end{equation}

There exists another solution to the vanishing of the boundary
variation terms that however will however 
involve rotated boundary conditions on
the $\sigma$ fields that do not agree with the NLSM limit. As always,
we begin with the fermions and choose 
$(\psi_{+i}+\eta e^{i\gamma} \psi_{-i})=0$ and then derive other
conditions by supersymmetric variation of the condition. The first
variation leads to eqn. (\ref{modbose}) which one can rewrite as two
separate conditions
\begin{eqnarray}
\left.\left(D_1\phi_i+\frac{1-e^{i\gamma}}{1+e^{i\gamma}}D_0\phi_i\right)
\right|_{x^1=0} =0 \\
\left.\frac{{\bar \sigma}-e^{i\gamma}\sigma}{1+e^{i\gamma}}\right|_{x^1=0}=0 \\
F_i|_{x^1=0}=0
\end{eqnarray}
The second equation is clearly in contradiction with the rotations
implied on $\sigma$ in the NLSM. Further supersymmetric variation
implies the following boundary condition on the twisted chiral
superfield (subject to $\theta^+ = \eta \theta^-$ and
$\overline{\theta}^+ = \eta \overline{\theta}^-$)
\begin{equation}
(\Sigma - e^{-i\gamma} \overline{\Sigma})|_{x^1=0}=0\quad.
\end{equation}
It is easy to verify that the boundary
terms in the ordinary and supersymmetric variations vanish.
This solution has also been recently proposed in \cite{HIV}. As emphasised 
earlier, this solution does not agree with the boundary conditions 
on the fields in the vector multiplet as derived in the NLSM limit.

The case of ``mixed'' boundary conditions (i.e., some fields have
Dirichlet boundary conditions imposed on them) also goes through.
The boundary contact term is again given by eqn. (\ref{boundterm}).
The boundary condition on the fields in the vector multiplet
are as in the all Neumann case and one can verify that 
all boundary terms in the ordinary and supersymmetric variations
vanish.

\subsection{Discussion}

	We would now like to discuss some of the implications of
the analysis of boundary conditions in the GLSM for the case of
the quintic at the Fermat point. In particular we would like to address
the question of whether we can identify the branes corresponding to our
various choices of boundary conditions. 
 
The choice of all Neumann boundary conditions on the matter fields is
clearly suggestive of a D6-brane. However we have to decide what
boundary conditions are appropriate for the $p$ field. It is useful to
consider for this purpose a slightly different example, without a
superpotential, namely that of the ${\cal O}(-3)$ line bundle over $\BP^2$.
In this case, clearly the construction of a D4-brane wrapping the $\BP^2$
would require that there be Dirichlet boundary conditions on the charge
$-3$ field. Otherwise we would not obtain a compact D4-brane. A similar
situation obtains in the quintic and hence we choose $p=0$ for the
D6-brane in the large volume. The $G=0$ condition is imposed on the
ground state by continuity from the bulk.  

Similarly we could identify the cases of the ``mixed''
Neumann-Dirichlet boundary conditions with lower-dimensional branes
while we always maintain Dirichlet boundary conditions on the $p$
field. However the charges of these lower-dimensional branes will not be
the minimum charge.  

However there could be other situations
where we impose Neumann boundary conditions that involve the $p$ field.
For instance we could impose conditions of the form $p\phi_1^5=c$ where
$c$ is a complex constant. Associated to this condition would be another
Neumann condition involving the $p$ field and the $\phi_1$. We can
then choose individual boundary conditions on the rest of the fields.
The interpretation of these boundary conditions would be different. 

Let us now consider the quintic at a point in its K\"ahler moduli
space where it admits a description as a Gepner model. B-type
boundary states arising from the Recknagel-Schomerus construction\cite{RS}
have been discussed in \cite{quintic}. The important point to note with
regard to the Recknagel-Schomerus construction is that the boundary
states arise from tensoring boundary states of individual minimal
models. Thus, they can only arise in our construction
by imposing boundary conditions on individual fields. 
We will for the moment focus our attention on the boundary
states labelled $|00000\rangle_B$ (There are five such states forming
a $\BZ_5$ orbit.). The analysis in \cite{quintic} shows that one
of these boundary states corresponds to the pure six-brane in the
large volume limit.

The Gepner point is in the LG phase (see section 2) 
where  we have seen that the only allowed boundary conditions
on individual fields are Dirichlet and hence all RS states
(including the one which carries pure six-brane charge)
must arise in this class. (See \cite{doug-diac} for a related
discusson.)  Consider now the all Neumann case
that we considered in the GLSM with $p=0$. 
The boundary condition on the bosonic fields as we have described
earlier are
(for $i=1,\ldots,5$)
$$
\left.\left\{D_1\phi_i+\frac{i\theta}{2\pi r}
D_0\phi_i+i\sqrt{2}\eta
\frac{{\bar \sigma}-e^{i\gamma}\sigma}{1+e^{i\gamma}}\phi_i
\right\}\right|_{x^1=0}=0 
$$
It is of interest to ask what happens to these boundary conditions
as $r \rightarrow 0$ while keeping $\theta$ fixed at some non-zero
value. In this limit $e^{i\gamma}\rightarrow -1$ and the above boundary
conditions tend to a Dirichlet boundary condition.
$$
D_0\phi_i+i\eta {{({\bar \sigma}+\sigma)}\over\sqrt{2}}\phi_i|_{x^1=0}=0 
$$
Interestingly, the $\sigma$ part in the above equation
comes out precisely of the form required because the fermionic
boundary condition becomes precisely the Dirichlet combination.
However quantum corrections (due to worldsheet
instantons) become important for small values of $r$ and further 
lines of marginal stability may be crossed in taking this limit. Though our
classical analysis may thus need to be modified, the above
result  is suggestively in agreement with the pure six-brane
appearing in the LG phase as an all Dirichlet state.
The limitations of our argument become clear once we consider
the cases of lower branes (such as the pure D4-brane) which do not
appear as a RS boundary state at the Gepner point.

We now comment on the extension of the results of this paper to other
examples. There are a number of straightforward extensions that require
very little beyond the techniques of this paper itself. 
All examples which are given by complete intersections of hypersurfaces
in weighted projective spaces or a tensor product of such spaces can be dealt
with. In such cases the analogues of the $p$-field are such that they
can be decoupled in the NLSM limit. Thus the rest of the analysis would
proceed much as in the single modulus case.

\section{Conclusion}
  In this paper we have taken the initial steps in what appears to 
be an useful programme of trying to understand, in a microscopic
description, D-branes in large domains
of the moduli space of the Calabi-Yau backgrounds in which they live.
The explicit nature of the boundary conditions that we impose on the
matter fields may be appropriately translated into the more general
characterisations of branes in the large-volume phase as zero sections
of the corresponding bundles on the CY manifold. Thus contact could be
made with other, more geometric techniques for understanding the
various properties of these branes in the large volume limit.  

The analysis of the GLSM in this paper
has been restricted to the open-string
channel. More information can be extracted by also investigating
the closed-string channel. Some related issues have already been 
considered in \cite{HV,HIV}. 

Another question which  needs to be
considered is the addition of (marginal) deformations corresponding to
gauge fields on the brane as well as its moduli including
the introduction of Chan-Paton factors.  This will involve describing
vector bundles on Calabi-Yau manifolds or their submanifolds. One
construction which easily fits the boundary GLSM is the use of monads
(as suggested in \cite{quintic}). It is
clear that this will involve techniques which appeared in the context of
$(0,2)$ versions of the closed string GLSM\cite{wittenphases,distler}
given that only half of the $(2,2)$ supersymmetry is preserved on the
boundary. For example, one has to introduce boundary fermions which
are sections of the appropriate vector bundle\cite{later}. The fermions $\nu^m$
which appeared in sec. 2.3.2 are objects of this type (they are sections
of the normal bundle).

A related problem
is the classification of boundary topological  observables in
the twisted version of the GLSM. This naturally leads to 
the next step in this program i.e., the use of the GLSM
description of D-branes to determine the superpotential in the
worldvolume theory of these branes. We believe that the results of this
paper are a useful step in proceeding towards this goal.  This 
(superpotential)
can be used as a check of mirror symmetry by computing the same in the
mirror manifold. Some aspects of these issues from a slightly
different viewpoint have been considered in\cite{vafamirror,ketal,kklm2}.
\bigskip

\noindent {\bf Acknowledgments} We would like to thank
Albion Lawrence and Hirosi Ooguri for collaboration during the initial 
stages of this work.
We are also indebted to them for numerous useful discussions and detailed
correspondence. We would also like to thank A.L. for a critical reading
of an earlier version of this paper and related correspondence. 
T.S. would like to thank S. Das and A. Sen for useful discussions. 
T.S. also acknowledges a Junior Research Fellowship of the 
Institute of Mathematical Sciences, Chennai where a part of this
work was done.  S.G. is supported in part by the Department of Science 
and Technology, India under the grant SP/S2/E-03/96.

\end{document}